\documentclass[conference]{IEEEtran}
\usepackage[left=0.62in,right=0.62in,top=0.75in,bottom=1in]{geometry}
\usepackage{graphicx}
\usepackage{cite}
\usepackage{amsmath}
\usepackage{amsfonts}
\usepackage{xcolor}
\usepackage{siunitx}
\usepackage{subcaption}
\usepackage{mathtools, cuted}
\usepackage{float}

\makeatletter
\def\ps@IEEEtitlepagestyle{%
  \def\@oddfoot{\mycopyrightnotice}%
  \def\@evenfoot{}%
}
\def\mycopyrightnotice{%
  {\footnotesize\hfill 978-1-7281-8510-1/20/\$31.00 \copyright2020 IEEE\hfill}%
  \gdef\mycopyrightnotice{}
}

\begin{document}
\title{Design of a Dynamic Parameter-Controlled Chaotic-PRNG in a 65 \textit{nm} CMOS process}


\author{
  \IEEEauthorblockN{
    Partha Sarathi Paul, Maisha Sadia, and Md Sakib Hasan
    \\
  }
  \IEEEauthorblockA{
    Department of Electrical and Computer Engineering
    \\
    University of Mississippi
    \\
    University, MS 38677, USA
    \\
    Email: ppaul@go.olemiss.edu, msadia@go.olemiss.edu, mhasan5@olemiss.edu
  }
}


%


\maketitle


\begin{abstract}
In this paper, we present the design of a new chaotic map circuit with a 65 $nm$ CMOS process. This chaotic map circuit uses a dynamic parameter-control topology and generates a wide chaotic range. We propose two designs of dynamic parameter-controlled chaotic map (DPCCM)-based pseudo-random number generators (PRNG). The randomness of the generated sequence is verified using three different statistical tests, namely, NIST SP 800-22 test, FIPS PUB 140-2 test, and Diehard test. Our first design offers a throughput of 200 MS/s with an on-chip area of 0.024 $mm\textsuperscript{2}$ and a power consumption of 2.33 $mW$. The throughput of our second design is 300 MS/s with an area consumption of 0.132 $mm\textsuperscript{2}$ and power consumption of 2.14 $mW$. The wider chaotic range and lower-overhead, offered by our designs, can be highly suitable for various applications such as, logic obfuscation, chaos-based cryptography, re-configurable random number generation, and hard-ware security for resource-constrained edge devices like IoT.

\end{abstract}


\section{Introduction}
\label{intro}
In recent decades, the nonlinear dynamic system has attracted the attention of researchers in the field of physics, biology, economics, finance, and engineering. Chaotic system is a particular type of nonlinear dynamic system whose behavior is characterized by deterministic equations. These irregular (aperiodic) but deterministic systems are very sensitive to the initial states. Even a tiny difference in the initial state makes the trajectory of two chaotic sequences, from the same chaotic system, significantly diverged. This phenomenon is popularly known as the butterfly effect. Some significant properties including, deterministic-aperiodicity and initial-state-sensitivity make the chaotic system very popular in security applications, including pseudo-random number generator (PRNG), cryptography, image encryption, and logic obfuscation. However, the present-day advancement in machine-learning algorithms is making the chaos-based security systems vulnerable to potential adversaries. Therefore, the chaos-based security community remains in an ongoing  quest for a superior-quality chaotic system. A higher dimensional chaotic system promises a more complex and secure chaotic sequence but brings the computation-cost into play as well. Hence, the design of a new chaotic system, with a simple structure but improved performance, is vital.

In this paper, we present the design of a new chaotic map circuit with a 65 $nm$ CMOS process. The general framework of the map circuit includes the three-transistor chaotic map circuit, proposed by Dudek \textit{et al.} \cite{dudek2003compact}, and dynamic parameter-control topology, proposed by Hua \textit{et al.} \cite{hua2015dynamic}. We propose the design of a novel PRNG circuit using our dynamic parameter-controlled chaotic map (DPCCM). Three statistical tests, NIST, FIPS, and Diehard, are performed on the PRNG output to verify the randomness of the generated sequence.

The remaining portion of the paper is organized as follows: section \ref{circuitdesign} describes the structure of the DPCCM. The proposed PRNG design and statistical analysis are described in section \ref{PRNG_design} and \ref{test}, respectively. Section \ref{overhead} talks about the overhead analysis of the proposed PRNG. Finally, section \ref{conclusion} gives the concluding remarks.

\section{Design of DPCCM}
\label{circuitdesign}
The transistor-level design of the DPCCM circuit is done using a 65 $nm$ CMOS process in Cadence, with the supply voltage of 1.2 $V$. The structure of the circuit can be described by separating it into three abstraction levels: a) the map circuit, b) the chaotic oscillator, and finally, c) the DPCCM.
\subsection{Map circuit}
The schematic of the three-transistor map circuit is shown in Fig. \ref{map}(a). The size of the transistors, $M_{1}$, $M_{2}$, and $M_{3}$, are carefully chosen to get a V-shaped transfer characteristic, which is vital for generating a chaotic sequence \cite{shanta2018design}. The gate voltage of $M_{2}$ (i.e. $V_{c}$) is used as the control voltage of the map.

\subsection{Chaotic oscillator}
The chaotic oscillator (shown in Fig. \ref{map}(b)) comprises of two feed-back connected map circuits, using the same control voltage. The initial state voltage, $X_n$, is applied through the clock, $clk_{ini}$. $X_n$ passes through the first map circuit block and we get the first output voltage, $V_{out1}$. $V_{out1}$ is then passed to the map circuit block in the feedback path through another clock, $clk_{a}$. The first iteration loop concludes when the generated output voltage, $V_{out2}$, is passed to the first map circuit through the clock, $nclk_{a}$. As the cycle continues, we get two analog voltages per iteration. Since, the clock signals, $clk_{a}$ and $nclk_{a}$, are non-overlapping, we use them to sample-and-hold the analog output voltage of the chaotic oscillator. Conventional designs use capacitors in the sample-and-hold circuit to store the data. In this topology, to reduce the on-chip area overhead, the parasitic capacitance of the pass transistors are leveraged to store the data. The total area consumption of the chaotic oscillator, along with the clocking circuits, is 0.556 $\mu m\textsuperscript{2}$ and the total power consumption is 18.4 $\mu W$. Fig. \ref{bifurcation}(a) shows the bifurcation plot of the chaotic oscillator and Fig. \ref{LE}(a) shows the corresponding Lyapunov exponent.

\subsection{DPCCM}
\begin{figure}
\centering
\includegraphics[scale=0.30]{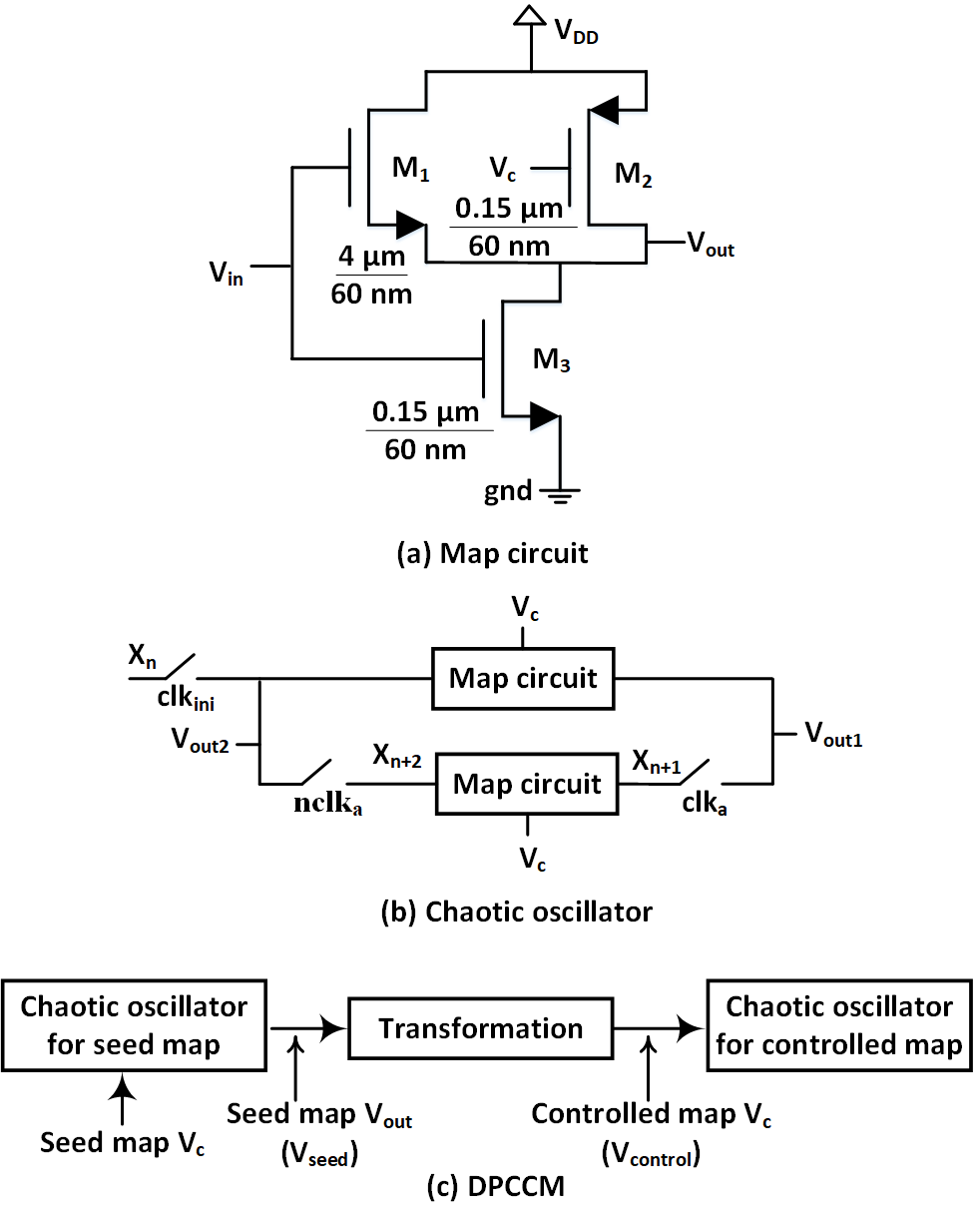}
\caption{\small{The schematic of DPCCM}}
\label{map}
\end{figure}

The block diagram of the DPCCM is presented in Fig. \ref{map}(c). The analog output voltage from a chaotic oscillator (seed map) undergoes a linear transformation before going into another chaotic oscillator (controlled map) as its control voltage parameter. The linear transformation is chosen in such a way that the output voltage of the first chaotic map is always mapped into a range of control voltages that generates chaotic sequence from the controlled map. The transformation is done according to Eq (\ref{transformation})\cite{hua2015dynamic}.
 \begin{equation}
    V_{control}=\frac{(b_{max}-b_{min})}{(a_{max}-a_{min})}\times(a_{max}-V_{seed})+b_{min} 
    \label{transformation}
\end{equation} 
In Eq (\ref{transformation}), $V_{seed}$ denotes any particular output voltage from the seed map, while $V_{control}$ is the transformed control voltage for the controlled map. The chaotic range for the control voltage is within $a_{max}$ and $a_{min}$ whereas, the span of the seed map output voltage is from $b_{max}$ to $b_{min}$.

\begin{figure}
\centering
  \subcaptionbox{\small{Chaotic oscillator}\label{fig3:a}}{\includegraphics[width=1.6 in]{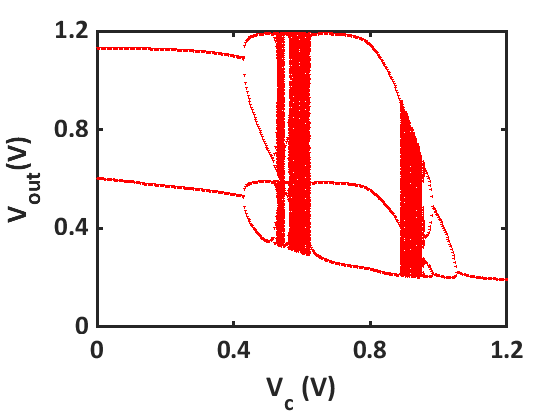}} \hspace{1em}%
  \subcaptionbox{\small{DPCCM}\label{fig3:b}}{\includegraphics[width=1.6 in]{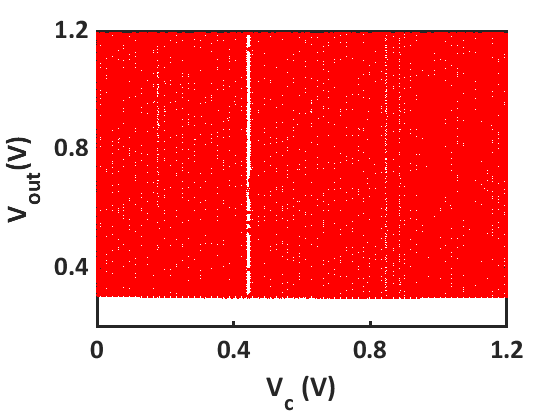}}%
  \caption{\small{(a) Bifurcation plot of the chaotic oscillator. The x-axis shows the common control voltage parameter for both feed-back connected  map circuits. (b) Bifurcation plot of the DPCCM. The x-axis shows the control voltage parameter of the seed map. The y-axis represents the steady-state output voltage in both plots.}}
  \label{bifurcation}
\end{figure}

\begin{figure}
\centering
  \subcaptionbox{\small{Chaotic oscillator}\label{fig3:a}}{\includegraphics[width=1.6in]{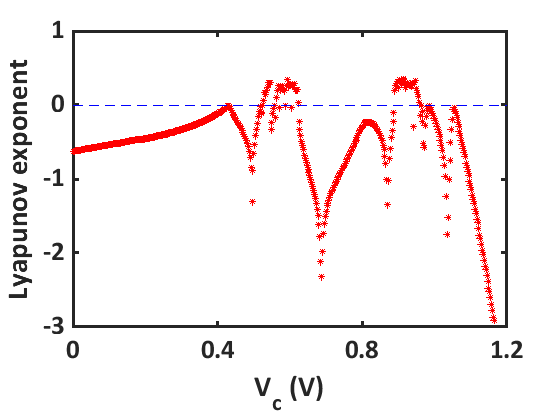}}\hspace{1em}%
  \subcaptionbox{\small{DPCCM}\label{fig3:b}}{\includegraphics[width=1.6in]{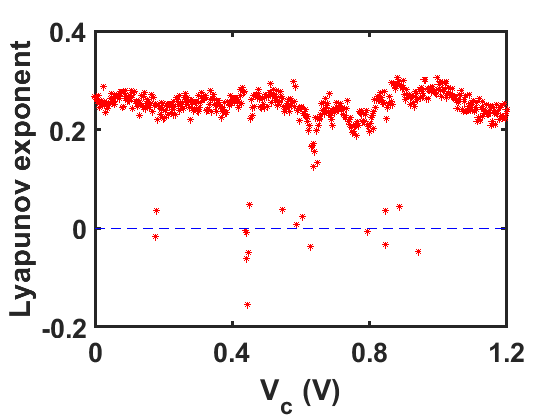}}%
  \caption{\small{(a) The plot of Lyapunov exponent of the chaotic oscillator. The x-axis shows the common control voltage parameter for both feed-back connected map circuits. (b) The plot of Lyapunov exponent of the DPCCM. The x-axis shows the control voltage parameter of the seed map. The y-axis, in both plots, represents the calculated Lyapunov exponent values from the steady-state output voltage. Positive values of Lyapunov exponent indicate chaotic operation.}}
  \label{LE}
\end{figure}

The term, dynamic parameter-control, comes from the fact that, at each iteration, the control voltage of the controlled map is being controlled by the output of the seed map. It is obvious from Fig. \ref{bifurcation}(a) and Fig. \ref{LE}(a) that, there are two distinct chaotic regions in the chaotic oscillator's output map, which shows positive Lyapunov exponent. The first range of the control voltage is between 0.5275 $V$ and 0.6225 $V$. The second range is between 0.89 $V$ to 0.9525 $V$. The output voltage covers a larger range for the case of the first control-voltage range (0.5275 $V$ - 0.6225 $V$). Moreover, it is found that the transistor delay increases with the increase of the control-voltage \cite{shanta2019design}. Hence, we chose the transformation of the DPCCM within the range of the first control voltage. The seed map and the controlled map starts with the same initial voltage. The bifurcation diagram from the DPCCM and the corresponding Lyapunov exponent is shown in Fig. \ref{bifurcation}(b) and Fig. \ref{LE}(b), respectively. The bifurcation plot of the DPCCM shows a wider chaotic range, compared to the bifurcation plot of the chaotic oscillator. A chaotic map circuit with a wider chaotic range can be extremely useful in applications like chaos-based logic and side channel obfuscation and re-configurable chaotic pseudo-random number generators (PRNG), where a wider chaotic range offers a larger design space. Here, we show the application of our DPCCM topology in a PRNG circuit.

\section{Design of PRNG}
\label{PRNG_design}

\begin{figure}
\centering
\includegraphics[scale=0.30]{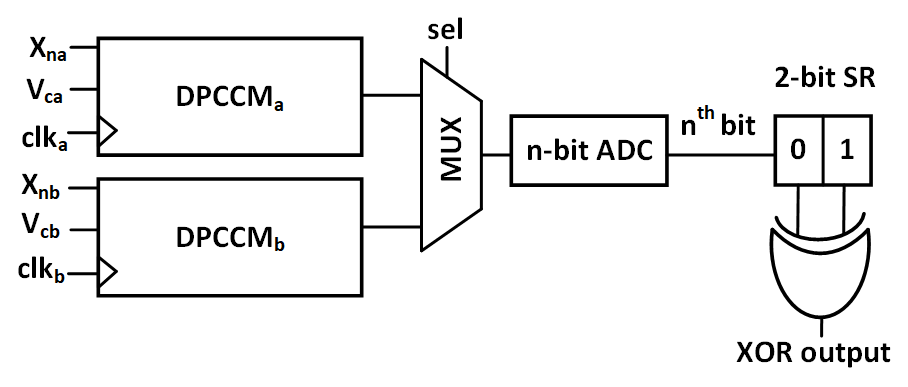}
\caption{\small{Pseudo-random number generator circuit}}
\label{PRNG}
\end{figure}

The DPCCM topology is implemented to design a PRNG. The schematic of the proposed PRNG is shown in Fig. \ref{PRNG}. Two DPCCM blocks, $DPCCM_{a}$ and $DPCCM_{b}$, are used in the PRNG design to get the required entropy in the generated sequence. Two phase-shifted clocks, $clk_{a}$ and $clk_{b}$, runs the DPCCM blocks, $DPCCM_{a}$ and $DPCCM_{b}$, respectively. As the clocks are phase-shifted, the selector pin, $sel$, of the multiplexer (MUX) can choose one output at a time and pass the output to an n-bit analog to digital converter (ADC). The analog voltages, generated at the even number of iterations, are sampled from DPCCM blocks. The ADC converts each analog voltage to an n-bit binary and stores the n\textsuperscript{th} bit into a 2-bit shift register. As soon as two binary bits are available from two DPCCM blocks, the data is XORed and sent to the output pin.  Two DPCCM blocks start the iteration with two unequal initial voltages, $X_{na}$ and $X_{nb}$. Each DPCCM block runs 2 million iteration loops. Therefore, we get a sequence of 1 million '1's and '0's at the XOR output. The seed map control voltages of DPCCM blocks, $V_{ca}$ and $V_{cb}$ are kept the same. The positive Lyapunov exponent, in the first chaotic range, of the chaotic oscillator (shown in Fig. \ref{LE}(a)) peaks for a control voltage value of 0.5925 $V$. Hence, the seed map control voltage is set to 0.5925 $V$. One hundred unique initial voltage pairs are used to generate a sequence containing 100 million binary bits, where each initial voltage pair corresponds to 1 million bits. The randomness in this generate sequence is tested in three different statistical test suites, NIST, FIPS, and Diehard test.

In addition to the above-mentioned PRNG design, we also propose a second PRNG design. In design-I, we take the n\textsuperscript{th} bit of an n-bit ADC whereas, in design-II, we capture the last 3 bits of an n-bit ADC. There are three 2-bit shift registers and three 2-input XOR gates in design-II. Each DPCCM block provides 3 binary bits for every even iteration. Therefore, 6 binary bits from the two DPCCM blocks are stored in those three 2-bit shift registers. When all six bits are available, they are bit position-wise XORed. That means, the n\textsuperscript{th}-bit of $DPCCM_{a}$ is XORed with n\textsuperscript{th} bit of  $DPCCM_{b}$. Thus, after running 2 million iteration cycles, we get 3 million-bit sequences from design-II. The throughput of design-II is three-times higher than design-I. For the statistical tests, a sequence of 100 million bits is generated from design-II as well.

\section{Statistical testes results}
\label{test}
\subsection{NIST SP 800-22 Test Suite}
 National Institute of Standards and Technology (NIST) test suite offers 15 statistical tests to measure the randomness in a sequence \cite{rukhin2001statistical}. The test was performed with a bit-stream length of 1 million. The significance level was set to 0.01. That means, a sequence with 100 million bits will pass a particular test if at least 96 out of the 100 bit-streams generate a $p-value$ of greater than 0.01. The 100 million-bit sequence, generated from design-I, passes all 15 tests with an ADC of bit-size $\geq$ 8. Whereas, design-II passes the test with an ADC of bit-size $\geq$ 10. Table \ref{nist_table} shows the NIST test result for design-I (8-bit ADC) and design-II (10 bit ADC).

 \begin{table}
  \centering
  \caption{NIST Test results for design-I and design-II\\
  (*shows the average of multiple tests)}
\scalebox{0.9}{
  \renewcommand{\arraystretch}{1.2}
\begin{tabular}{|c|c|c|}
\hline 
\textbf{NIST TEST}
& \multicolumn{2}{c|}{\textbf{Pass rate}}\tabularnewline
\cline{2-3} \cline{3-3} 
 & \textbf{design-I} & \textbf{design-II}\tabularnewline
\cline{2-3} \cline{3-3} 
 & 8-bit ADC & 10-bit ADC\tabularnewline
\hline 
\hline 
\textbf{Frequency}  & 0.99 & 1.00\tabularnewline
\hline 
\textbf{Block frequency} & 0.98 & 0.99\tabularnewline
\hline 
\textbf{Cumulative sums*} & 1.00 & 1.00\tabularnewline
\hline 
\textbf{Runs} & 1.00 & 1.00\tabularnewline
\hline 
\textbf{Longest runs of ones} & 0.99 & 0.98\tabularnewline
\hline 
\textbf{Rank} & 1.00 & 0.99\tabularnewline
\hline 
\textbf{FFT} & 0.99 & 0.99\tabularnewline
\hline 
\textbf{Non-overlapping template*} & 0.99 & 0.99 \tabularnewline
\hline 
\textbf{Overlapping template }& 0.99 & 0.98\tabularnewline
\hline 
\textbf{Universal} & 1.00 & 1.00\tabularnewline
\hline 
\textbf{Approximate entropy} & 0.98 & 0.99\tabularnewline
\hline 
\textbf{Random excursion*} &0.99  & 0.99 \tabularnewline
\hline 
\textbf{Random excursion variant*} &1.00  & 1.00\tabularnewline
\hline 
\textbf{Serial*} & 1.00 & 0.99\tabularnewline
\hline 
\textbf{Linear complexity} & 1.00 & 1.00\tabularnewline
\hline 
\end{tabular}}
\label{nist_table}
\end{table}

\subsection{FIPS PUB 140-2}
Federal Information Processing Standards Publications (FIPS PUB) 140-2 Test was developed by NIST \cite{pub2001140}. FIPS tests the randomness of a binary sequence by dividing it into 20,000-bit blocks. The blocks are subjected to 4 sub-tests - the Monobit test, Poker test, Runs test and Long run test. The Monobit test counts the number of 1's in the 20,000-bit block. This number must be within the range of (9725,10275) to pass the test. The Poker test divides the 20,000-bit block into 5,000 successive 4-bit segments. The 4-bit segment can have 16 possible values and the occurrence of these 16 values is counted and stored. This sub-test verifies the uniformity of the 4-bit segment. Runs test counts and stores the maximum sequence of consecutive bits of 1's or 0's in a 20,000-bit block. A run of 26 or more of either 1's or 0's is defined as a Long run. The total number of Long runs in a 20,000-bit block is counted in the total failure. Table \ref{fips_table} shows the FIPS test result for design-I and design-II. FIPS divides the 100 million sequence into 5,000 blocks of 20,000-bits. The second column of Table \ref{fips_table} shows the number of blocks passing the test and the last 4 columns show the number of failed blocks under corresponding sub-tests.

\begin{table}
  \centering
  \caption{FIPS test results}
\scalebox{0.9}{
  \renewcommand{\arraystretch}{1.2}
  
\begin{tabular}{|c|c|c|c|c|c|}
\hline 
PRNG & \textbf{Total success} & Monobit & Poker & Runs & Long run\tabularnewline
\hline 
\hline 
design-I & \textbf{4999} & - & - & 1 & -\tabularnewline
\hline 
design-II & \textbf{4996} & - & 3 & - & 1\tabularnewline
\hline 
\end{tabular}}
\label{fips_table}
\end{table}

\subsection{Diehard Statistical Test Suite}

\begin{figure}
\centering
  \subcaptionbox{\small{design-I (8-bit ADC)} \label{fig3:a}}{\includegraphics[width=1.6 in]{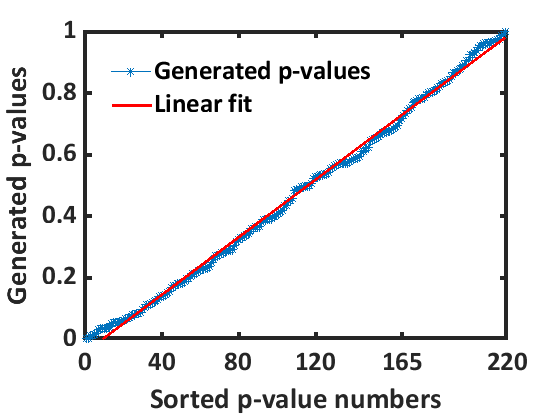}} \hspace{1em}%
  \subcaptionbox{\small{design-II (10-bit ADC)}\label{fig3:b}}{\includegraphics[width=1.6 in]{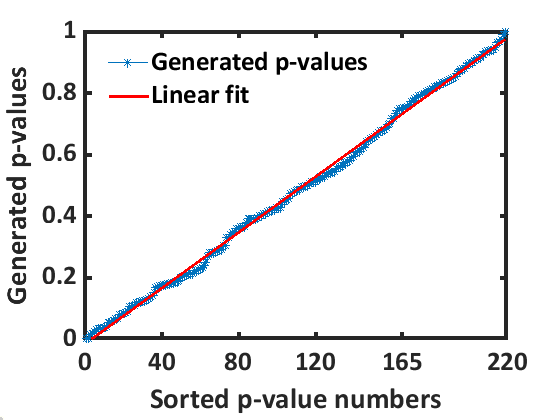}}%
  \caption{\small{Diehard statistical test results}}
  \label{diehard_Plot}
\end{figure}

Diehard Statistical Test Suite, developed by George Marsaglia, offers 15 sub-tests, that generate 219 $p-values$ \cite{diehard_Marsaglia}. The sequence is considered to be random if the $p-values$ are within the range of [0,1). A sequence will fail the test if it produces six or more (out of 219) $p-values$ of either 0 or 1. Our test sequence size is 100,000,032 bits. Fig.  \ref{diehard_Plot} shows the plot of the generated $p-values$, sorted in ascending order. The linear fit shows close conformity with the generated $p-value$ trend and demonstrates the randomness of our sequence.

\section{Overhead analysis of the PRNG}
\label{overhead}
Each of the two DPCCM blocks, shown in Fig. \ref{PRNG}, used in the design, consumes an area of 1.1 $\mu m\textsuperscript{2}$ and a power of 36.8 $\mu W$. The 8-bit ADC, used in design-I, can be implemented from the proposed design of Wei \textit{et al.} in \cite{wei20110}. This 8-bit ADC takes 0.024 $mm\textsuperscript{2}$ on-chip area and consumes 1.8 $mW$ power. As we can see, the ADC is responsible for most of the area and power consumption of the circuit making it the primary overhead contributor to the whole PRNG design. The maximum bit generation rate of this 8-bit ADC is 250 MS/s. However, design-I generates binary sequence with a rate of 200 MS/s. The throughput of the slower component dictates the overall throughput of the system. Hence, the throughput of the overall PRNG system, with design-I, is limited to 200 MS/s. We have determined that our design-II can offer a maximum throughput of 600 MS/s (i.e. three times higher than design-I). The 10-bit ADC, used in design-II, can be implemented from the design reported by Ma \textit{et al.} \cite{ma2011low}. This 10-bit ADC covers an on-chip area of 0.132 $mm\textsuperscript{2}$ and consumes 1.6 $mW$ power. For our design-II, we are taking the last three ADC bits at every even iteration. Therefore, the throughput from the 10-bit ADC, with 100 MS/s bit-rate,  will eventually become 300 MS/s (i.e. three times 100 MS/s) in our design. For this reason, the throughput of the overall PRNG system, with design-II, is limited to 300 MS/s. The total area overhead for design-I, including chaotic oscillators, ADC, shift register, MUX, and XOR gate, is 0.024 $mm\textsuperscript{2}$ and the power overhead is 2.33 $mW$. For design-II, total area and power overheads are 0.132 $mm\textsuperscript{2}$ and 2.14 $mW$, respectively. Table \ref{overhead_comp} presents a comparison between the PRNG designs of this paper and prior works on different PRNG circuits. Compared to previous works, both of our designs cover significantly less on-chip area. Our design-I is a suitable choice for applications with high area-constraint whereas design-II offers better throughput. It should be noted that design-II can be generalized for higher-bit ADC while compromising the area and power.

\begin{table}
\label{overhead_comp}
  \centering
  \caption{Overhead comparison}
  
\scalebox{0.85}{
  \renewcommand{\arraystretch}{1.2}
  
\begin{tabular}{|c|c|c|c|c|c|c|}
\hline 
Parameter & \multicolumn{4}{c|}{Reported works} & \multicolumn{2}{c|}{This work}\tabularnewline
\cline{2-7} \cline{3-7} \cline{4-7} \cline{5-7} \cline{6-7} \cline{7-7} 
 & \cite{pareschi2010implementation}  & {\cite{li2011period}}  & {\cite{yang2004chaos}}  & {\cite{shanta2019design}}  & design-I & design-II\tabularnewline
\hline 
\hline 
Technology ($nm$)  & 180  & 180  & 180  & 65  & \textbf{65} & \textbf{65}\tabularnewline
\hline 
Supply voltage ($V$)  & 1.8  & 1.8  & 1.8  & 1.2  & \textbf{1.2} & \textbf{1.2}\tabularnewline
\hline 
Area ($mm\textsuperscript{2}$)  & 0.126  & 0.275  & 0.767  & 0.132  & \textbf{0.024} & \textbf{0.132}\tabularnewline
\hline 
Power ($mW$)  & 22  & 13.9  & 37  & 2.12  & \textbf{2.33} & \textbf{2.14}\tabularnewline
\hline 
Throughput ($MS/s$)  & 100  & 6400  & 120  & 100  & \textbf{200} & \textbf{300} \tabularnewline
\hline 
\end{tabular}}
\label{overhead_comp}
\end{table}

\section{Conclusion}
\label{conclusion}

A new design of a chaotic map circuit, with dynamic parameter control, is presented in a 65 $nm$ CMOS process. It has been shown that this design provides a wide chaotic region in the bifurcation plot, offering a large design space for various security applications such as, logic obfuscation, chaos-based cryptography and re-configurable random number generator circuits. The map circuit is used in a novel PRNG design. The randomness of the generated sequence is verified using three statistical tests. It is found that the dynamic parameter-control leads to a low-overhead PRNG design. Future work may include, the exploration of various CMOS topologies and different post-processing schemes for the bit generation.

\bibliographystyle{IEEEtran}

\bibliography{main.bib}




\end{document}